\documentclass[12pt]{article}

\usepackage{amsmath}
\usepackage{amssymb}
\usepackage{amsthm}
\usepackage{amsbsy}
\usepackage{amsfonts}
\usepackage{dsfont}
\usepackage{color}
\usepackage{graphicx}
\usepackage{alltt}

\newcommand\BibTeX{{\rmfamily B\kern-.05em \textsc{i\kern-.025em b}\kern-.08em
T\kern-.1667em\lower.7ex\hbox{E}\kern-.125emX}}

\newcommand{\bm}[1]{{\boldsymbol {#1}}}

\newcommand{\bA}{{\bm A}}

\newcommand{\bC}{{\bm C}}
\newcommand{\bc}{{\bm c}}

\newcommand{\bx}{{\bm x}}
\newcommand{\by}{{\bm y}}

\newcommand{\bS}{{\bm S}}

\newcommand{\bz}{{\bm z}}

\newcommand{\bzero}{{\bm 0}}
\newcommand{\bone}{{\bm 1}}

\newcommand{\bmu}{\bm\mu}

\newcommand{\bEta}{\bm\eta}

\newcommand{\btheta}{\bm\theta}

\newcommand{\bLambda}{\bm\Lambda}

\makeatletter
\let\@fnsymbol\@arabic
\makeatother

\begin{document}

\title{Model-based dose finding under model uncertainty using general
  parametric models} \author{Jos\'{e} Pinheiro\footnote{Janssen
    Research \& Development, Raritan, NJ, 08869, USA.}, Bj\"{o}rn
  Bornkamp\footnote{Novartis Pharma AG, CH-4002 Basel, Switzerland.},
  Ekkehard Glimm$^2$ and Frank Bretz$^2$ }

\date{}

\maketitle

\begin{abstract}

  Statistical methodology for the design and analysis of clinical
  Phase II dose response studies, with related software
  implementation, are well developed for the case of a normally
  distributed, homoscedastic response considered for a single
  timepoint in parallel group study designs.  In practice, however,
  binary, count, or time-to-event endpoints are often used, typically
  measured repeatedly over time and sometimes in more complex settings
  like crossover study designs.  In this paper we develop an
  overarching methodology to perform efficient multiple comparisons
  and modeling for dose finding, under uncertainty about the
  dose-response shape, using general parametric models.  The framework
  described here is quite general and covers dose finding using
  generalized non-linear models, linear and non-linear mixed effects
  models, Cox proportional hazards (PH) models, etc.  In addition to
  the core framework, we also develop a general purpose methodology to
  fit dose response data in a computationally and statistically
  efficient way.  Several examples, using a variety of different
  statistical models, illustrate the breadth of applicability of the
  results.  For the analyses we developed the \texttt{R} add-on
  package \texttt{DoseFinding}, which provides a convenient interface
  to the general approach adopted here. % \texttt{SAS} code is available
  % by the authors upon request.
\end{abstract}

\section{Introduction}
\label{sec:introduction}

Finding the right dose is a critical step in pharmaceutical drug
development.  The problem of selecting the right dose or dose range
occurs at almost every stage throughout the process of developing a
new drug, such as in microarray studies \cite{lin:2012}, in-vitro
toxicological assays \cite{bret:hoth:2003}, animal carcinogenicity
studies \cite{hoel:port:1994}, Phase I studies to estimate the maximum
tolerated dose \cite{neue:bran:gspo:2008}, Phase II studies covering
dose ranging and dose-exposure-response modeling \cite{hsu:2009},
Phase III studies for confirmatory dose selection, and post-marketing
studies to further explore dose response in specific subgroups defined
by region, age, disease severity and other covariates
\cite{rube:1995a,rube:1995b,iche4:1994}.  In recent years,
considerable effort has been spent on improving the efficiency of dose
finding throughout drug development
\cite{ting:2006,chev:2006,kris:2006}.  Despite these efforts, however,
improper dose selection for confirmatory studies, due to lack of
sufficient dose response knowledge for both efficacy and safety at the
end of Phase II, remains a key driver of the ongoing pipeline problem
experienced by the pharmaceutical industry
\cite{phrma:2007,pinh:2010}.

Statistical analysis methods for late development dose finding studies
can be roughly categorized into modelling approaches to characterize
the dose response relationship \cite{pinh:bret:bran:2006,thom:2006}
and multiple test procedures for dose response signal detection
\cite{stew:rube:2000} or confirmatory dose selection
\cite{tamh:hoch:dunn:1996,bret:maur:bran:2009}.  Hybrid approaches
combine aspects of multiple testing with modeling techniques to
overcome the shortcomings of either approach
\cite{tuke:cimi:heys:1985, bret:pinh:bran:2005}.  More recently,
considerable research has focused on extending these methods to
response-adaptive designs that offer efficient ways to learn about the
dose response through repeated looks at the data during an ongoing
study
\cite{drag:2010,drag:hsua:padm:2007,jone:layt:rich:2011,born:bret:dett:2011}.

Most statistical methodology for dose response analysis has been
introduced in the context of a normally distributed, homoscedastic
endpoint, with a parallel group design, in which each patient receives
only one treatment. In practice, however, one often faces more complex
study designs (e.g., cross-over designs), where a heteroscedastic or
non-normally distributed endpoint is measured (e.g., binary, count and
sometimes time-to-event data).  One approach is to extend the existing
methodology using generalized non-linear models or generalized
non-linear mixed effects models. However, these extensions are
typically specific to each new situation. In addition general purpose
software for these types of models is not available and a case-by-case
implementation requires a major coding effort. In this paper, we
describe an overarching hybrid approach, combing multiple comparisons
and modeling, to the analysis of dose response data for general
parametric models and general study designs, that allows for a
straightforward computer implementation.

\begin{figure} [h!]
  \begin{center}
  \includegraphics[width=0.75\textwidth]{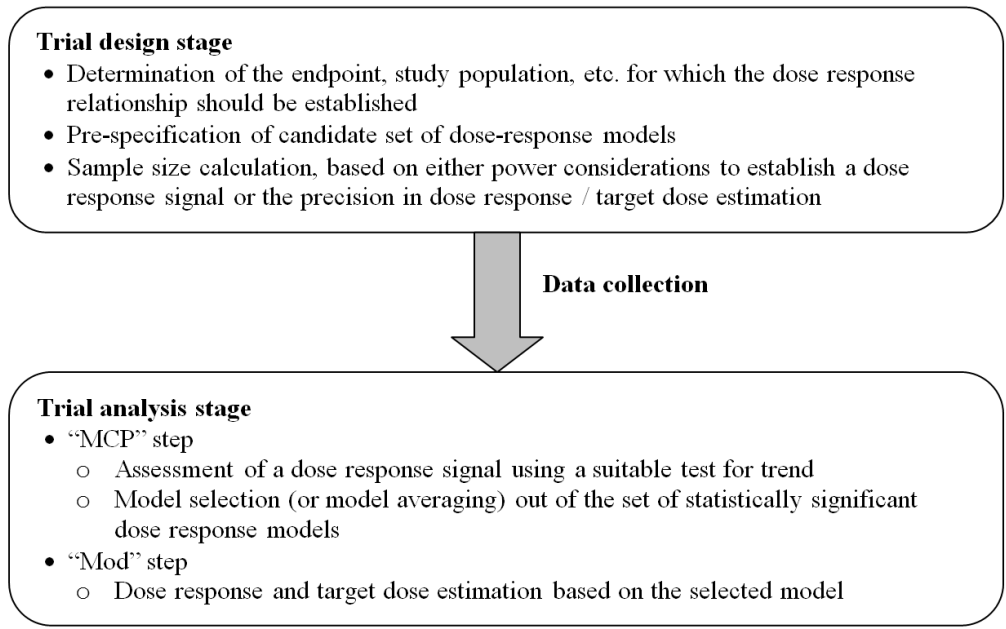}
  \end{center}
\caption{Overview of MCPMod approach}
\label{fig:mcpmod}
\end{figure}

More specifically, we extend the MCPMod approach
\cite{bret:pinh:bran:2005}, which was originally introduced for
normal, homoscedastic, independent data.  This approach provides the
flexibility of modeling for dose response and target dose estimation,
while accounting for model uncertainty through the use of multiple
comparison and model selection/averaging procedures. The approach can
be described in two main steps (Fig.~\ref{fig:mcpmod}).  At the trial
design stage the clinical team needs to decide on the core aspects of
the trial design, as in any other trial. Specific for MCPMod is that a
candidate set of plausible dose response models is pre-specified at
this stage, based on available pharmacokinetic data/dose response
information from similar compounds, etc. This gives rise to a set of
optimal contrasts used to test for the presence of a dose response
signal consistent with the corresponding candidate models. In case of
large model uncertainty, this candidate set should be chosen to cover
a sufficiently diverse set of plausible dose response shapes

The trial analysis stage consists of two main steps: The \textit{MCP}
and the \textit{Mod} steps. The \textit{MCP} step focuses on
establishing evidence of a drug effect across the doses,
i.e., detecting a statistically significant dose response signal for
the clinical endpoint and patient population investigated in the
study. This step will typically be performed using an efficient test
for trend, adjusting for the fact that multiple candidate dose
response models are being considered. If a statistically significant
dose response signal is established, one proceeds with
determining a reference set of significant dose response models by
discarding the non-significant models from the initial candidate
set. Out of this reference set, a best model is selected for dose
estimation in the last stage of the procedure, using existing
non-linear regression methods \cite{bate:watts:1988}. The selected
dose-response model is then employed to estimate target doses using
inverse regression techniques and possibly incorporating information
on clinically relevant effects. The precision of the estimated doses
can be assessed using, for example, bootstrap methods.

The original MCPMod method proposed by \cite{bret:pinh:bran:2005} was
intended to be used with parallel group designs with a normally
distributed, homocedastic response. Although that covers a good range
of dose finding designs utilized in practice, the restrictions of the
original method create serious practical limitations to its wider use
in drug development. For example, binary, count, and time-to-event
endpoints, though frequently used in many disease indications, are not
covered by the original MCPMod formulation. Likewise, longitudinal
patient data, like in crossover studies, with its implicit
within-patient correlation, cannot be properly analyzed with the
original formulation of the MCPMod methodology.  In what follows, we
extend the MCPMod methodology, to perform dose response modeling and
testing in the context of general parametric models and for general
study designs, in a similar way as \cite{hoth:bret:west:2008} extended
certain simultaneous inference procedures.  Note that, even though we
focus on extending the MCPMod approach, the results of this paper
remain valid, in particular, if only a multiple comparison or a
modeling approach is to be applied.

We introduce the proposed extension in Section~\ref{sec:gen}.  In
Section~\ref{sec:nldr} we describe an efficient approach for dose
response model fitting, which is evaluated in terms of asymptotic
considerations and simulations. In Section~\ref{sec:appl} we use
several case studies to illustrate the proposed methodology and its
implementation with the \texttt{R} add-on package \texttt{DoseFinding}
\cite{DoseFinding}.  In summary the method is illustrated for binary,
overdispersed count, time-to-event data (based on the Cox PH model)
and longitudinal data with patient specific random effects.

\section{Generalized MCPMod}
\label{sec:gen}

This section describes an extension of the original MCPMod approach
that can be used whenever the response variable can be described by a
parametric model in which one of the parameters captures the dose
response relationship. We show how the basic ideas and concepts of the
original MCPMod can be extended to this setting.

\subsection{Basic concepts, notation and assumptions}
\label{ssec:basic}
In the original description of MCPMod, the expected value of the
response is utilized as the parameter capturing the dose response
relationship. The key idea of the extended version of MCPMod is to
decouple the dose response model from the expected response, focusing,
instead, on a more general characterization of dose response via some
appropriate parameter in the probability distribution of the
response. To be more concrete, let $\by$ denote the response vector of
an experimental unit in the trial (e.g., a patient) which has been
assigned a dose $x$. The following results can easily be extended to
the case of multiple doses $\bx$, if needed. We assume that $\by$
follows the distribution function given by
\begin{equation}
  \label{eq:gMod}
  \by \sim F\left(\bz,\bEta,\mu(x)\right),
\end{equation}
where $\mu(x)$ denotes the dose response parameter, $\bEta$ the vector
of nuisance parameters, and $\bz$ the vector of possible
covariates. The key features of extended version of MCPMod can then be
formulated with respect to $\mu(x)$, such as:
\begin{itemize}
\item accounting for uncertainty in the dose response model via a set
  of candidate dose response models,
\item testing of dose response signal via contrasts based on dose response shapes,
\item model selection via information criteria, or model averaging to
  combine different models, and
\item dose response estimation and dose selection via modeling.
\end{itemize}

Because all dose response information is assumed to be captured by
$\mu(x)$, the interpretability of this parameter is critical for
communicating with clinical teams, choosing candidate dose response
shapes, specifying clinically relevant effects for target dose
estimation, etc. To better illustrate this point, consider a
time-to-event endpoint that is assumed to follow a Weibull
distribution. The Weibull distribution is typically parameterized by a
scale parameter $\lambda$ and shape parameter $\alpha$, neither of
which has an easy clinical interpretation. For the purpose of MCPMod
modelling the model could be reparameterized in terms of the median
time to event $\mu = \left[\log(2)\right]^{1/\alpha}/\lambda$ and
$\alpha$, and then one would use $\mu(x)$ as an interpretable dose
response parameter.

All dose response models of interest in this paper can be expressed as
\begin{equation}
  \label{eq:dose responsemod}
  f(x,\btheta) = \theta_0 + \theta_1f^0(x,\btheta^0),
\end{equation}
where $f^0$ denotes the so-called standardized model function and
$\btheta^0$ its parameter vector. For example, for $f^0(x,\theta^0)=x$
one obtains the linear model $ f(x,\btheta) = \theta_0 + \theta_1 x$
and for $f^0(x,\theta^0)=x/(x+\theta^0)$ the Emax dose response model
$ f(x,\btheta) = \theta_0 + \theta_1 x/(x+\theta^0)$; see
~\cite{bret:pinh:bran:2005, born:pinh:bret:2009} for more examples.
Dose response models of the form~(\ref{eq:dose responsemod}) are
specified as candidate models for $\mu(x)$. Covariates may also be
included in~(\ref{eq:dose responsemod}), at the model fitting stage,
but we leave them out for now to keep the notation simple.

Assume that $K$ distinct doses $x_1,\ldots,x_K$ are utilized in a trial,
with $x_1$ denoting placebo. Assume further that $M$ candidate models
$f_1,\ldots,f_M$ have been chosen to capture the model uncertainty
about $\mu(x)$. We define the dose response parameter vector associated to
candidate model $m$ as $\bmu_m =
\left(\mu_{m,1},\ldots,\mu_{m,K}\right),$ where
$\mu_{m,i}=f_m(x_i,\btheta), i=1,\ldots,K, m=1,\ldots,M$.

For the purpose of obtaining estimates of the dose response, we
initially consider an analysis-of-variance (ANOVA) parametrization for
the dose response parameter $\mu(x_i) = \mu_i, \; i = 1,
\ldots,K$. That is, we allow a separate parameter $\mu_i$ to represent
the dose response at each dose level. Let $\widehat{\bmu}$ denote the
vector of estimated dose response parameters under the ANOVA
parametrization, obtained using the appropriate estimation method for
the general parametric model~(\ref{eq:gMod}) via maximum likelihood
(ML), generalized estimating equations (GEE), partial likelihood,
etc. Note that these type of ANOVA estimates are easily available from
standard statistical software packages. The key assumption
underpinning the extended version of MCPMod is that $\widehat{\bmu}$
has an approximate distribution ${\cal N}\left(\bmu, \bS\right),$
where $\bS$ denotes the variance-covariance of $\widehat{\bmu}$. This
assumption can be shown to hold for most parametric estimation
problems, such as, generalized linear models, parametric time-to-event
models, mixed-effects models, GEE, etc. Note that $\bS$ is a function
of $n$ and converges to $\mathbf{0}$ as $n\rightarrow \infty$.
Furthermore, $\bm S$ may or may not depend on $\bm x$. For example,
dependence on $\bm x$ arises if unequal variances for different dose
levels $x_i$ are considered.  The estimation of $\btheta$ is done in a
separate second stage based on $\widehat{\bmu}$ and an estimate
$\widehat{\bm S}$ of its covariance matrix.  Section $\ref{sec:nldr}$
explains this in detail.

\subsection{Implementation of the MCP step}
\label{ssec:impl}
The \textit{MCP} step consists of specifying a set of candidate
models for the dose response relationship $\mu(x)$. To that end, one needs to
specify families of candidate models (\textit{e.g.}, linear, Emax,
logistic, quadratic). In addition, to derive optimal model contrasts,
one needs to determine \textit{guesstimates} for the non-linear parameters $\bm
\theta^0$, such as the ED$_{50}$ parameter for the Emax model.
Note that the shape of the dose response model function is
determined by the parameter $\bm \theta^0$, which is why only
guesstimates for this parameter are needed to derive optimal model
contrasts. As mentioned earlier, the clinical interpretability of
$\mu(x)$ is critical for this step. Further details on and strategies
for the specification of candidate models and corresponding
guesstimates are given in \cite{pinh:born:bret:2006}.

Given these guesstimates, each candidate model shape determines an
optimal contrast for a trend test to evaluate the associated dose
response model signal, such as a linear trend or a trend based on an
Emax model with ED$_{50}=2$.  The optimal contrasts are applied to the
previously described ANOVA estimates $\widehat{\bmu}$, with the
associated asymptotic distribution used for implementing the
corresponding tests (\textit{i.e.}, critical values and p-values). It
can be shown that the (optimal) contrast for testing the hypothesis of
a flat dose-response profile with maximal power for a single candidate
model shape $\bmu_m$ is given by
\begin{equation}
  \label{eq:optCont}
  \bc^{opt} \propto \bS^{-1}\left(\bmu^0_m - \frac{\bmu^{0\prime}_m \bS^{-1}
      \bone}{\bone^\prime \bS^{-1} \bone}\right),
\end{equation}
where $\bmu^0_m=(f^0_m(x_1,\btheta^0),\ldots,f^0_m(x_K,\btheta^0))'$
and $\btheta^0$ is the parameter for which guesstimates are required,
see \cite{bret:pinh:bran:2005} and Appendix A.
For convenience, we normalize the contrast coefficients such that
$||\bc_{opt}|| = 1$.

The implementation of contrast tests for the candidate models is done
similarly to the original MCPMod approach. Let
$\bc_1^{opt},\ldots,\bc_{M}^{opt}$ represent the optimal contrasts
corresponding to the set of candidate models and
$\bC^{opt}=\left[\bc_1^{opt}\cdots\bc_{M}^{opt}\right]$ the associated
optimal contrast matrix. The contrast estimates are then given by
$\left(\bC^{opt}\right)^{\prime} \widehat{\bmu}$, being asymptotically
normally distributed with mean $\left(\bC^{opt}\right)^{\prime} \bmu$
and covariance matrix $\left(\bC^{opt}\right)^{\prime}\bS \bC^{opt}.$
It follows that the asymptotic z-test statistic for the m$^{th}$
candidate model hypotheses $H_0:\left(\bc_m^{opt}\right)^{\prime}\bmu
= 0$ vs. $H_0:\left(\bc_m^{opt}\right)^{\prime}\bmu > 0$ is given by
$z_m =
\left(\bc_m^{opt}\right)^{\prime}\widehat{\bmu}/\left[\left(\bC^{opt}\right)^{\prime}
  \bS \bC^{opt} \right]_{m,m}^{1/2},$ with $[\bA]_{m,m}$ denoting the
m$^{th}$ diagonal element of the matrix $\bA$. The test statistic used
for establishing an overall dose response signal is the maximum
$z_{(M)} = \max_m z_m$ of the individual model test
statistics. Critical values for tests with (asymptotically) exact
level $\alpha$ can be derived from the joint distribution of $\bz =
\left(z_1,\ldots,z_M\right)$, which is easily obtained from the joint
distribution of the contrast estimates given previously, and using
  \begin{equation}
    \label{eq:maxZ}
    P(z_{(M)}>q) = 1 - P(z_{(M)} \leq q) = 1 - P(\bz \leq q \bone).
  \end{equation}
Multiplicity adjusted p-values for the individual model contrast tests
can be derived similarly. The \texttt{mvtnorm} package in \textsf{R}
includes functions to calculate quantiles and probabilities for the
underlying multivariate normal distributions \cite{genz:bret:2009}.

If the optimal contrasts and the critical values also depend on $\bS$,
one needs guesstimates for nuisance parameters appearing in the
covariance matrix at the planning stage, as well. This is a difference
compared to the normal homoscedastic setting without covariates, where
$\bS$ is proportional to a diagonal matrix with the the reciprocal of
the group sample sizes on the diagonal. Once data becomes available,
the z-statistics for the model contrast tests are calculated and their
maximum used for the dose response test. At this stage, one can obtain
the estimated $\widehat{\bS}$ matrix from the observed data, and use
this to recalculate optimal contrasts and the critical value for the
test.  Note that the guesstimates for the parameters $\bm \theta^0$
are not recalculated based on the observed data, as this would lead to
a serious Type I error rate inflation. For the purpose of the multiple
contrast test, the guesstimates pre-specified at the planning stage
for $\bm \theta^0$ should be used.

\subsection{Implementation of the Mod step}
\label{sec:implmod}

Once a dose response signal is established, one proceeds to the
\textit{Mod} step, fitting the dose response profile and estimating
target doses based on the models identified in the \textit{MCP}
step. There are many ways to fit the dose response models to the
observed data, including approaches based on maximizing the likelihood
(ML) or the restricted likelihood.  However, a direct ML approach
requires the derivation of the likelihood in every specific case, with
a considerable amount of model-specific coding involved. We therefore
suggest an alternative two-stage approach to dose response model
fitting that utilizes generalized least squares. This approach is
described in more detail in Section~\ref{sec:nldr}. It relies on
asymptotic results, but has the appeal of being of general purpose
application, as it depends only on $\widehat{\bmu}$ and
$\widehat{\bS}$. In addition, as shown later in the simulation study,
its finite and large sample properties are similar to those of the
approaches based on the full likelihood.

Model selection can be based on the maximum z-statistics, or using
information criteria, such as the AIC or the BIC. Estimates of the
latter under the model fitting approach are discussed in the next
section. Estimation of target doses is done based on the selected
fitted model for the dose response parameter
\cite{born:pinh:bret:2009}.

Alternatively, model averaging approaches can be used to avoid the
need to select a single model. The individual AIC and BIC values for
the candidate models with significant contrast test statistics
determine the model averaging weights \cite{buck:burn:augu:1997}. This
applies both to dose response and target dose estimation.

The generalization of MCPMod described in this section has focused on
testing and estimation associated with the full dose response profile,
that is, including the response at placebo and the entire dose range
utilized in the study. In practice, there are cases in which inference
might focus on the placebo-adjusted dose response (or more generally a
control-adjusted response), that is, the dose response with the
placebo or control effect subtracted $ f_C(x,\btheta) = f(x,\btheta) -
f(0, \btheta)$. This could become relevant, for example, if covariates
are added to the placebo response $\theta_0$ in~(\ref{eq:dose
  responsemod}).  In the context of time-to-event data, the focus on
placebo-adjusted dose response will occur naturally when modeling the
hazard ratio as a function of dose. As shown in
Appendix~\ref{ref:diff2plac}, all results presented in this section,
including the derivation of optimal contrasts, as well as the model
fitting results described in the next section, apply equally in the
context of placebo-adjusted dose response.

\section{Non-linear dose response model fitting using a two-stage,
  generalized least squares approach}
\label{sec:nldr}

In this section we describe an efficient methodology for fitting
nonlinear dose response models that can be used for the \textit{Mod}
step of the MCPMod methodology. The fitting is done in two stages:
First, the ANOVA estimates $\widehat{\bm \mu}$ and $\widehat{\bm S}$
introduced in Section~\ref{sec:gen} are obtained. Second, the
parameter $\bm \theta$ is estimated by fitting the dose response model
to the ANOVA estimates from the first step using a generalized least
squares estimation criterion.  In Section~\ref{sec:asymp}, we
establish consistency and asymptotic normality of this estimate. In
Section~\ref{sec:simul} we assess the accuracy of the asymptotic
results via a simulation study.

\subsection{Asymptotic results}
\label{sec:asymp}

Assume that we have dose response estimates $\widehat{\bm \mu} =
(\widehat{\mu}_1, \ldots, \widehat{\mu}_K)'$ obtained from an
ANOVA-type parameterization of $\bmu$ which allows for unrelated mean
responses at each of the $K$ dose levels; see
Section~\ref{ssec:basic}.  These estimates are assumed to be
asymptotically multivariate normal distributed with a covariance
matrix consistently estimated by $\widehat{\bm S}$.  The estimates
$\widehat{\bm \mu}$ and $\widehat{\bm S}$ are easily available from
standard statistical packages, see Section~\ref{sec:appl} for examples
using \texttt{R}.  Next, we fit the non-linear dose response model
$f(x,\bm \theta)$ to the estimates $\widehat{\bm \mu}$ by minimizing
the generalized least squares criterion
\begin{equation}
\label{eqn:psi0}
\widehat{\bm \Psi}(\bm \theta)=(\widehat{\bm \mu}-\bm f(\bm x, \bm \theta))'
{\bm A_n}(\widehat{\bm \mu}-\bm f(\bm x, \bm \theta))
\end{equation}
with respect to $\bm \theta$ to obtain the estimates $\widehat{\bm
  \theta}$. In Equation~\eqref{eqn:psi0} we have $\bm f(\bm x, \bm
\theta)=(f(x_1, \bm \theta), \ldots, f(x_K, \bm \theta))'$ and $\bm
A_n$ denotes a symmetric positive definite matrix. We assume that $\bm
A_n \overset{P}{\rightarrow} \bm A$, where $\overset{P}{\rightarrow}$
denotes convergence in probability. In practice we will always use
$\bm A_n = \widehat{\bm S}^{-1}$, but this would unnecessarily
restrict the discussion at this stage.

Let $\bm \theta_0$ denote the true value of the parameter $\bm
\theta$.  In Appendix \ref{app:b} we show that, under mild regularity
conditions, $\widehat{\bm \theta}$ is a consistent estimator of $\bm
\theta_0$, i.e., $\widehat{\bm \theta} \overset{P}{\rightarrow} \bm
\theta_0$.  Furthermore, we have the asymptotic multivariate normality
\begin{equation}
\label{eq:asymptnorm}
\sqrt{a_n}(\widehat{\bm \theta}-\bm
\theta_0)\overset{d}{\rightarrow}N(\bm 0,\bm B(\bm \theta_0)'\bm
M(\bm \theta_0)\bm B(\bm \theta_0)),
\end{equation}
where $\bm M(\bm \theta)=a_n\bm F(\bm \theta)'\bm A\bm S \bm A\bm F(\bm
\theta)$ and $\bm B(\bm \theta)=(\bm F(\bm \theta)'\bm A\bm F(\bm
\theta))^{-1}$, $\bm F(\bm \theta)$ denotes the $d \times k$ matrix of
partial derivatives $\frac{d f(x_i,\bm \theta)}{d \theta_h}$
($i=1,...,k$,$\quad h=1,...,d$), $a_n$ a non-decreasing sequence of
values increasing to infinity as the sample size $n$ goes to infinity.

Selecting $\bm A_n=\bm S^{-1}$ would be the best choice, if $\bm S$
were known. Provided that $a_n \bm S \overset{P}{\rightarrow} \bm
\Sigma$ as $n\rightarrow \infty$, the previous formulas in this case
simplify to $\sqrt{a_n}(\widehat{\bm \theta}-\bm
\theta_0)\overset{d}{\rightarrow}N(\bm 0, (\bm F(\bm \theta_0){\bm
  \Sigma}^{-1}\bm F(\bm \theta_0)')^{-1})$ with $\widehat{\bm \theta}$
minimizing
\begin{equation}
\label{eqn:psi}
\widehat{\bm \Psi}(\bm \theta)=(\widehat{\bm \mu}-\bm f(\bm x, \bm \theta))'
{\bm S}^{-1}(\widehat{\bm \mu}-\bm f(\bm x, \bm
\theta))
\end{equation}
with respect to $\bm \theta$. Because $\bm S$ is not known, we will
typically use a consistent estimate $\widehat{\bm S}$ of it in
(\ref{eqn:psi}). If the assumptions about the covariance matrix are
wrong in the sense that $a_n \widehat{\bm S}$ does not converge to
$\bm \Sigma$, then $\widehat{\bm \theta}$ will still be consistent and
have the asymptotic normal distribution given in
(\ref{eq:asymptnorm}). In this regard, the suggested estimator is
similar to Huber's robust estimator (\cite{hube:1967}; Section 4.6 in
\cite{digg:heag:lian:1994}), but this aspect will not be further
utilized.

Note that this two-stage, generalized least squares estimate is quite
similar to the ML estimate: For normal homoscedastic data both
approaches lead to exactly the same estimate, while, for example, in
generalized linear model settings the two estimators have the same
large-sample variance; see Chapter 4.3 in \cite{dobs:2002}.

The computational advantage of using this two-stage approach is that
the target function in \eqref{eqn:psi} that is optimized numerically
is low-dimensional: The dimension is equal to the number of different
dose levels and the target function can thus be evaluated quite
efficiently, while the target function in a full likelihood approach
depends on the complete data set.  This difference in speed becomes
relevant in clinical trial settings, as often extensive clinical trial
simulations are used to evaluate proposed study designs. Another
advantage of the proposed method is its broad applicability to general
parametric models.

Model selection criteria are generally defined as $-2\log(L)+\dim(\bm
\theta)\tau$, where $L$ denotes the likelihood function evaluated at
the maximum likelihood estimate and $\tau$ a penalty for the number of
parameters, which depends on the model selection criterion. One
approach to model selection is hence to use $\widehat{\bm
  \Psi}(\widehat{\bm \theta})+\dim(\bm \theta)\tau$ to compare
different dose response models fitted based on the same $\widehat{\bm
  \mu}$ and $\bm S$. This criterion is motivated by the fact that, for
normally distributed homoscedastic data without covariates, these two
approaches are equivalent in terms of selecting the same model: The
likelihood function can be split into the sum of the deviation between
the observed data and $\widehat{\bm \mu}$ and the deviation between
$\widehat{\bm \mu}$ and $\bm f(\bm x, \widehat{\bm \theta})$. The
deviation of the individual data and $\widehat{\bm \mu}$ is identical
across the different dose response models, so that the criterion only
varies with the deviation between $\widehat{\bm \mu}$ and $\bm f(\bm
x, \widehat{\bm \theta})$, which, in case of normal data, is equal to
$\widehat{\bm \Psi}(\widehat{\bm \theta})$. In situations beyond the
normal case both approaches might lead to slightly different results,
however as (\ref{eqn:psi}) is roughly proportional to
$-2\cdot$log-likelihood of the ML estimate of $\bm \theta$, when
discarding the contribution of the first stage ANOVA-type fit (which
is equal for all dose response models considered) typically both
approaches will lead to very similar results. Subsequently $\tau=2$
will be used and we refer to this criterion as gAIC.

In the next subsection we investigate the properties of the proposed
asymptotic approximations via simulations before illustrating it
with several applications in Section~\ref{sec:appl}.

\subsection{Simulations}
\label{sec:simul}

In this section we evaluate the asymptotic performance of the
approximations provided in Section \ref{sec:asymp} for fitting a
single nonlinear dose response model. More specifically, we compare
the proposed methods with more traditional maximum likelihood
estimation by evaluating the dose response estimation accuracy using
simulations.  In addition we assess the coverage probability of the
resulting confidence intervals for the model parameter $\bm \theta$.

\subsubsection{Design of simulation study}
\label{sssec:SimulDes}

Throughout the simulations we assume five active dose levels 0.05,
0.2, 0.5, 0.8, 1 plus placebo. We investigate equal group sample sizes
of 15, 30, 50, 100, 300 and 1000 patients for each dose. The lower
range of the investigated sample sizes is realistic for typical Phase
II dose response studies, while the sample sizes of 300 and 1000 are
included to assess the asymptotic behavior.

\begin{figure} [h!]
  \begin{center}
  \includegraphics[width=0.75\textwidth]{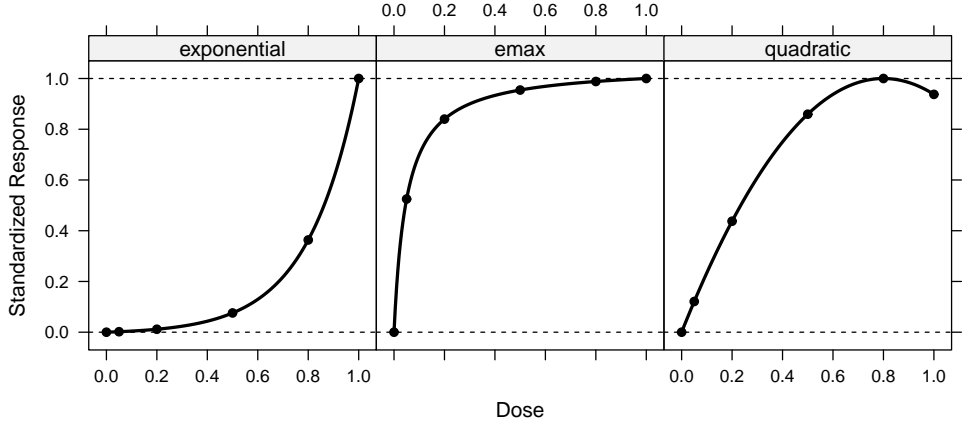}
  \end{center}
\caption{Dose Response Models used for simulation}
\label{fig:simmod}
\end{figure}

We investigate three types of data: binary data, overdispersed count
data using a negative binomial model and time-to-event data using a
Cox PH model for estimation.  Regarding dose response models, we will
utilize an Emax model $\mu(x,\theta)=\theta_0+\theta_1x/(\theta_2+x)$,
a quadratic model $\mu(x, \theta)=\theta_0+\theta_1x+\theta_2x^2$ and
an exponential model $\mu(x,
\theta)=\theta_0+\theta_1(\exp(x/\theta_2)-1)$.  In the simulations,
we set $\theta_2= 0.05$ for the Emax model, $\theta_2= 0.2$ for the
exponential model and $\theta_1/\theta_2=-5/8$ for the quadratic
model; see Figure \ref{fig:simmod} for the underlying model shapes.
The remaining parameters $\theta_0$ and $\theta_1$ are chosen such
that the power for testing the dose with the maximum treatment effect
against placebo at the 5\% one-sided significance level is 80\% for 30
patients per group.  This ensures a realistic range of sample sizes
(in terms of the signal to noise ratio) is investigated in the
simulations.

\begin{table}[h!]
\centering
\begin{tabular}{|l|l|l|l|}
\hline\noalign{\smallskip}
Data type & Quadratic & Emax & Exponential \\ \hline
binary        & $(-1.734,4.335,-2.7094)$ & $(-1.734,1.8207,0.05)$ &  $(-1.734,0.01176,0.2)$ \\
count         & $(2,-2,1.25)$ & $(2,-0.84,0.05)$ &  $(2,-0.005427,0.2)$\\
time-to-event      & $(0,-1.8876,1.1797)$ & $(0,-0.7928,0.05)$  & $(0,-0.005122,0.2)$ \\
normal      & $(0, 2.61, 1.633)$ & $(0, 1.097, 0.05)$  & $(0,0.007089, 0.2)$ \\
\noalign{\smallskip}\hline
\end{tabular}
  \caption{Dose Response Model Parameters $(\theta_0,\theta_1,\theta_2)$ used for simulation}
\label{tab:simmod}
\end{table}

Table \ref{tab:simmod} summarizes the three dose response model
specifications for each data type.  For binary data, Table
\ref{tab:simmod} gives the the mean on the logit scale.  For count
data, the logarithm of the mean is as specified in Table
\ref{tab:simmod} and the overdispersion parameter is 1. For
time-to-event data, we use an exponential distribution for data
generation with the log-means specified in Table \ref{tab:simmod} and
where observations larger than 10 are censored. The mean in the
placebo group is 1, so that the log-mean is equal to the difference in
log-hazard rates. The Cox PH model is formulated relative to the
control group, so that, in this case, the placebo parameter is set to
0 when estimating the dose response model. In addition, we include
normally distributed data with $\sigma = 1$ as a benchmark comparison,
since in this case the two-stage, generalized least squares (GLS) and
ML estimates coincide.

We use the two-stage approach from Section~\ref{sec:gen} to obtain (i)
an ANOVA-type model fit to the data by using either a generalized
linear model (binary and count data) or a Cox PH model (time-to-event
data) with ``dose'' treated as a factor and (ii) a dose response model
fit to the resulting dose response estimates obtained via generalized
least squares (\ref{eqn:psi}), together with the asymptotic results
from Section~\ref{sec:asymp}. In the simulations, we compare this
approach to nonlinear maximum likelihood (binary and count data) and
maximum partial likelihood (time-to-event data) estimation using the
same link functions as above. For the model fitting step, we assume
lower and upper bounds for the $\theta_2$ parameter of 0.001 and 5 for
the Emax and 0.05 and 5 for the exponential dose response model.

In addition, we assess the coverage probability for three different
methods of constructing confidence intervals for the dose response
model parameter $\bm \theta$.  First, we use the generalized least
squares (\ref{eqn:psi}) together with the asymptotic results from
Section~\ref{sec:asymp} (denoted below as GLS).  Second, we use
parametric bootstrap confidence intervals by sampling from the
multivariate normal distribution underlying the first stage ANOVA-type
estimates and then fitting the nonlinear model to each of these
samples using the GLS criterion from (\ref{eqn:psi}). The bootstrap
confidence intervals are then calculated by taking the 5\% and 95\%
quantiles of the observed sample. For each simulation we used 500
bootstrap samples (denoted below as GLS-B).  Finally, we use the
maximum likelihood fits and calculate confidence intervals based on
the inverse of the Hessian matrix and the usual asymptotic normality
assumptions (denoted below as ML).

\subsubsection{Results of simulation study}
\label{sssec:SimulRes}

Simulations were run with 2000 replications, using the
\texttt{DoseFinding} package version 0.9-1. To illustrate the
performance of the GLS and ML methods with regard to dose response
estimation, we calculated the root mean squared estimation error
averaged over the available doses
$\sqrt{\frac{1}{6}\underset{x\in\mathcal{D}}{\sum}(f(x,\widehat{\bm
    \theta})-f(x,\bm \theta))^2}$, where
$\mathcal{D}=\{0,0.05,0.2,0.6,0.8,1\}$.  Figures
\ref{fig:drest},~\ref{fig:estbin} and \ref{fig:est} in the Appendix
display the results.  It is evident from these plots that both
approaches can hardly be distinguished in terms of the mean squared
error, indicating that, in terms of dose response estimation, both
methods perform almost identically, even for small sample sizes.

\begin{figure}
  \begin{center}
    \includegraphics[width=0.7\textwidth]{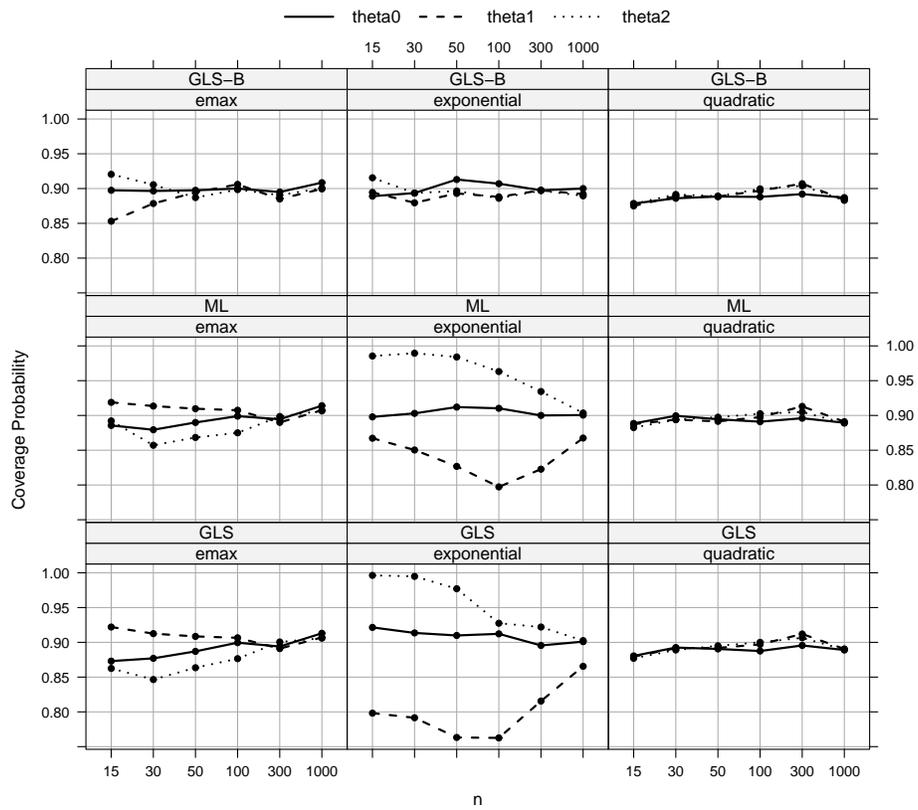}
  \end{center}
\caption{Empirical coverage probability (based on 2000 simulations),
  of 90\% confidence intervals.}
\label{fig:CIcount}
\end{figure}

Next, we assess the coverage probability for the three different
methods described at the end of Section~\ref{sssec:SimulDes}.  Figure
\ref{fig:CIcount} displays the results only for the count data case,
because the results for the binary and time-to-event data are nearly
indentical.  We observe that the asymptotic confidence intervals for
the GLS and ML methods perform very similarly for all three models
under investigation (Emax, exponential, quadratic).  Both methods
achieve the nominal 90\% coverage probability fairly well for the
quadratic model, even for small sample sizes. For the Emax model, the
nominal coverage probability is achieved at roughly 50-100 patients
per group, whereas for the exponential model the coverage probability
is achieved only at very large sample sizes (the poor performance of
standard asymptotic confidence intervals for nonlinear regression
models even for moderate sample sizes is well-known, see for example,
\cite{pedd:hase:2005}).  The reason for the better performance under
the quadratic model is the fact that it is linear in the model
parameters. One reason for why the confidence intervals perform worse
for the exponential model than for the Emax model might be that the
dose design used in the simulations allows an easier identification of
the model parameters under the Emax model, because there are more dose
levels in the lower part of the dose range than the upper part.

In contrast, the parametric bootstrap approach GLS-B achieves the 90\%
nominal coverage probability fairly well for all three dose response
models, even at sample sizes as small as 30 patients per group. The
GLS-B method thus performs always at least as well as the GLS and ML
methods, and the general recommendation is to use this in case of
small sample sizes. The approach is computationally more expensive, as
it requires repeated fitting of the dose response models, but the
bootstrap-based on the GLS two-stage fitting is computionally much
more efficient than a traditional bootstrap approach based on ML: The
GLS two-stage approach only depends on the ANOVA type estimates and
not the individual observations, which makes evaluation of
$\widehat{\bm \Psi}(\widehat{\bm \theta})$ computationally much
cheaper than evaluation of the full likelihood function.

In summary, we conclude that both the GLS and ML methods perform
similarly under the different dose response shapes, sample sizes, and
data types investigated in the simulation study.  This conclusion
holds both for the coverage probabilities, as well as for the average
estimation error.  As mentioned previously, however, the GLS method is
very general and computationally more efficient, which facilitates the
usage of computationally expensive techniques such as the bootstrap
approach.

\section{Applications}
\label{sec:appl}

\subsection{Longitudinal modeling of neurodegenerative disease}
\label{ssec:longitEx}
This example refers to a Phase 2 clinical study of a new drug for a
neurodegenerative disease. The state of the disease is measured
through a functional scale, with smaller values corresponding to more
severe neurodeterioration. The goal of the drug is to reduce the rate
of disease progression, which is measured by the linear slope of the
functional scale over time.

The trial design includes placebo and four doses: 1, 3, 10, and 30 mg,
with balanced allocation of $50$ patients per arm. Patients are
followed up for one year, with measurements of the functional scale
being taken at baseline and every three months thereafter. The study
goals are to (i) test the dose-response signal, (ii) estimate the
dose-response and (iii) select a dose to be brought into the
confirmatory stage of the development program.

The functional scale response is assumed to be normally distributed
and, based on historical data, it is believed that the longitudinal
progression of the functional scale over the one year of follow up can
be modeled by a simple linear trend. We use this example to illustrate
the application of MCPMod in the context of mixed-effects models.

We consider a mixed-effects model representation for
the functional scale measurement $y_{ij}$ on patient $i$ at time
$t_{ij}$:
\begin{equation}
  \label{eq:LME}
  y_{ij} = \left(\beta_0+b_{0i}\right) +
  \left(\mu(x_i)+b_{1i}\right)t_{ij} + \epsilon_{ij}, \;
  \left[b_{0i},b_{1i}\right]^\prime {\sim}
  N\left(\bzero, \bLambda \right) \; \mathrm{and} \; \epsilon_{ij}
  {\sim} N\left(0, \sigma^2 \right),\mathrm{ all \; stoch. \; independent}.
\end{equation}
The DR parameter in this case is the linear slope of disease
progression $\mu(x)$. If $\mu(x)$ is represented by a linear function
of dose $x$, the model in (\ref{eq:LME}) is a linear mixed-effects (LME)
model, else it becomes a nonlinear mixed-effects (NLME) model. In
particular, under the ANOVA parametrization discussed in
Section~\ref{ssec:basic}, $\mu(x_i) = \mu_i$, (\ref{eq:LME}) is
an LME model with different slopes for each dose.

The research interest in this study focuses on the treatment effect on
the linear progression slope. At $t=1$ year this is numerically equal
to the average change from baseline, and thus easily interpretable. At
the planning stage of the trial, the following assumptions were agreed
with the clinical team for the purpose of design:
\begin{itemize}
\item Natural disease progression slope = -5 points per year.
\item Placebo effect = 0 (\emph{i.e.}, no change in natural progression).
\item Maximum improvement over placebo within dose range = 2 points
  increase in slope over placebo.
\item Target (clinically meaningful) effect = 1.4 points increase in
  slope over placebo.
\end{itemize}
Guesstimates for the variance-covariance parameters were obtained from
historical data: $\mathrm{var}\left(b_{0i}\right) = 100$,
$\mathrm{var}\left(b_{1i}\right) = 9$ $\mathrm{corr}\left(b_{0i},
  b_{1i}\right) = -0.5$, and $\mathrm{var}\left(\epsilon_{ij}\right) =
9$. Under these assumptions, it is easy to see that the covariance
matrix of the ANOVA-type estimate $\widehat{\bmu}$ of the slopes
$\bmu=\left(\mu_{1 mg}, \mu_{3 mg}, \mu_{10 mg}, \mu_{30 mg}\right)'$
is compound-symmetric. With these concrete guesstimates, the diagonal
element is $0.1451$ and the off-diagonal element $0.0092$.

\begin{figure}
  \begin{center}
  \includegraphics[width=\textwidth]{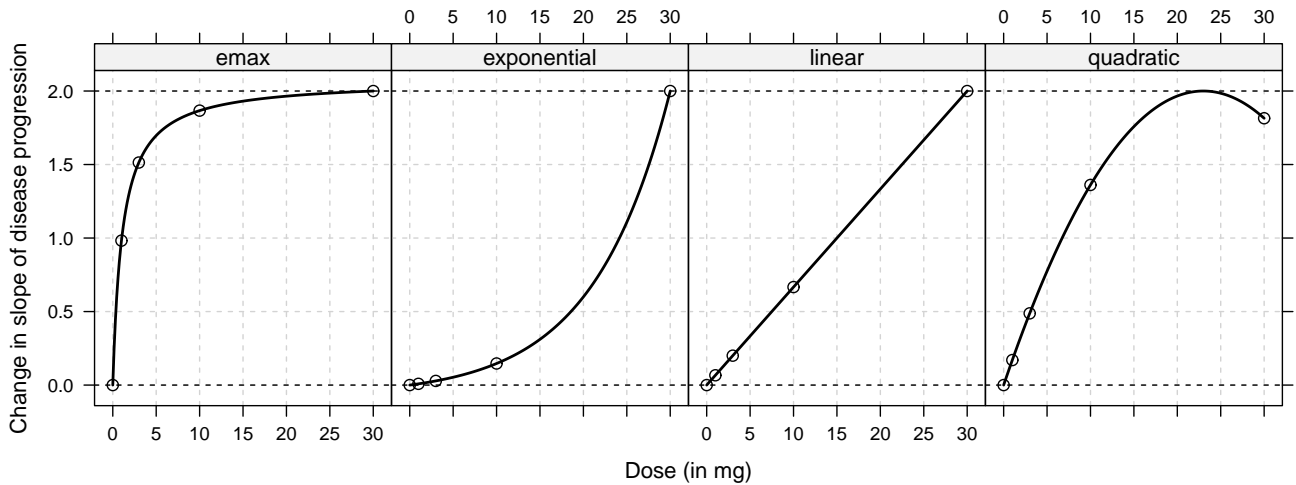}
  \end{center}
\caption{Candidate models for neurodenerative disease example.}
\label{fig:candModLong}
\end{figure}

From discussions with the clinical team, the four candidate models
displayed in Figure~\ref{fig:candModLong} were identified:
\begin{itemize}
\item Emax model with 90\% of the maximum effect at 10 mg,
  corresponding to an ED$_{50} = 1.11$
\item Quadratic model with maximum effect at 23 mg, corresponding to
  standardized model parameter $\delta = -0.022$
\item Exponential model with 30\% of the maximum effect occurring at
  20 mg, corresponding to a standardized model parameter $\delta =
  8.867$
\item Linear model
\end{itemize}

\begin{figure}
  \begin{center}
  \includegraphics[width=\textwidth]{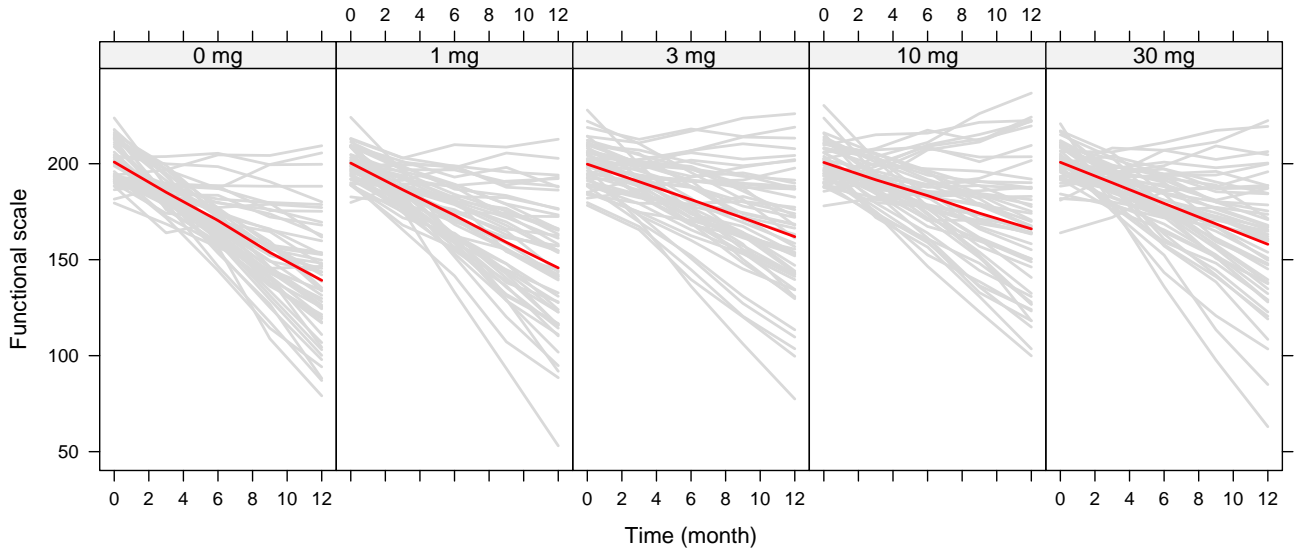}
  \end{center}
\caption{Simulated data for neurodegenerative disease example. Gray
  lines correspond to individual patient profiles, black line to loess
smoother.}
\label{fig:dataLong}
\end{figure}

For confidentiality reasons, the data from the actual trial cannot be
used here, so we utilize a simulated dataset with characteristics
similar to the original data with an Emax DR profile imposed on the
linear slopes $\mu(x)$. Figure~\ref{fig:dataLong} shows the simulated
data per dose, which is available in the \texttt{DoseFinding} package,
in the \texttt{neurodeg} data set.

In what follows we illustrate the individual steps of MCPMod along
with its implementation in \texttt{DoseFinding} package (version
0.9-6).

The $\widehat{\bmu}$ vector is estimated via an LME fit of data, which
can be done, for example, using the \texttt{lme} function in the
\texttt{nlme} R package, as illustrated below.
\begin{alltt}
data(neurodeg)
head(neurodeg, n=3)
>       resp id dose time
> 1 191.7016  1    0    0
> 2 178.3995  1    0    3
> 3 167.3385  1    0    6
fm <- lme(resp ~ as.factor(dose):time, neurodeg, ~time|id)
muH <- fixef(fm)[-1] # extract estimates
covH <- vcov(fm)[-1,-1]
\end{alltt}
The estimated slopes are $\widehat{\bmu} = \left(-5.099, -4.581,
  -3.220, -2.879, -3.520\right)^\prime$ with corresponding estimated
  variance-covariance matrix with compound symmetry structure with
  diagonal elements $0.149$ and off-diagonal elements $0.0094$.

  The optimal contrasts corresponding to the candidate models are
  calculated using the formula in~(\ref{eq:optCont}), with $\bS$ given
  by the estimated variance-covariance matrix of $\widehat{\bmu}$. The
  \texttt{DoseFinding} package includes the function \texttt{optContr}
  to calculate optimal contrasts
  based on~(\ref{eq:optCont}) as follows. 
\begin{alltt}
doses <- c(0, 1, 3, 10, 30)
mod <- Mods(emax = 1.11, quadratic=-0.022, exponential = 8.867, linear = NULL,
            doses = doses) # definition of candidate shapes
contMat <- optContr(mod, S=covH) # calculate optimal contrasts
\end{alltt}

The \texttt{MCTtest} function in the \texttt{DoseFinding} package
implements the optimal model contrast tests for $\widehat{\bmu}$ based
on the multiple comparison approach described in
Section~\ref{ssec:impl}. 
\begin{alltt}
MCTtest(doses, muH, S=covH, type = "general", critV = T, contMat=contMat)
> . . .
> Multiple Contrast Test:
>             t-Stat  adj-p
> emax         4.561 <0.001
> quadratic    3.680 <0.001
> linear       2.274 0.0249
> exponential  1.277 0.1818
> 
> Critical value: 2.275 (alpha = 0.025, one-sided)
\end{alltt}
The Emax, quadratic and linear model contrasts are all significant,
but the exponential model failed to reach significance. Therefore, the
significance of a dose-response signal is established and we can move
forward to estimating the dose-response profile and the target dose.

Two approaches can be used for model fitting in this example: the
two-stage GLS non-linear dose-response fitting method described in
Section~\ref{sec:nldr}, or mixed-effects modeling (linear and
nonlinear) incorporating a parametric dose response model for the
progression slope $\mu(x)$. We consider first the two-stage GLS
method, which is implemented in the \texttt{fitMod} function in
\texttt{DoseFinding}, illustrated in the call below for the Emax
model.
\begin{alltt}
fitMod(doses, muH, S=covH, model="emax", type = "general", bnds=c(0.1, 10))
> Dose Response Model
> 
> Model: emax 
> Fit-type: general 
> 
> Coefficients dose-response model
>     e0   eMax   ed50 
> -5.181  2.180  1.187 
\end{alltt}
The gAIC values (as discussed in Section \ref{sec:asymp})
corresponding to the fits for the Emax, quadratic, and linear models
are, respectively: 10.66, 11.07 and 24.22, indicating the better
adequacy of the Emax model. Note that the \texttt{DoseFinding} package
also includes an \texttt{MCPMod} function that performs
\texttt{MCTtest}, model selection and model fitting in one step.

The mixed-effects model fit approach in this case is illustrated below
for the Emax model using the \texttt{nlme} function
\begin{alltt}
nlme(resp ~ b0 + (e0 + eM * dose/(ed50 + dose))*time, neurodeg,
     fixed = b0 + e0 + eM + ed50 ~ 1, random = b0 + e0 ~ 1 | id,
     start = c(200, -4.6, 1.6, 3.2))
...
  Fixed: b0 + e0 + eM + ed50 ~ 1
        b0         e0         eM       ed50
200.451303  -5.178739   2.181037   1.198791
\end{alltt}
The estimated fixed effects from the NLME model are quite close to the
estimates obtained via the GLS two-stage method. The AIC values
corresponding to the Emax, quadratic and linear models under the
mixed-effects model fit are, respectively: 8352.60, 8353.10 and
8365.79 confirming the Emax as the best fitting model. It is
intriguing to see how similar the differences in AIC between the
different models are to the differences in gAIC values.

Estimates for the target dose, that is, the smallest dose producing an
effect greater than or equal to the target value of 1.4, can be
obtained with either of the model fitting approaches. The resulting
estimated target doses are 2.13 under the two-stage GLS method and
2.15 under the NLME model. Alternatively, model averaging methods
could have been used to estimate the target dose and the dose-response
profile.

\subsection{Binary and placebo-adjusted data}
\label{sec:df-pack}

In this section we will go through two concrete examples on how to use
the \texttt{DoseFinding} R package to apply MCPMod to binary data and
placebo-adjusted normal data. Only the required R commands are given
here, but not the output.

\subsubsection{Binary endpoint}
\label{sec:example}

This example is based on trial NCT00712725 from
\texttt{clinicaltrials.gov}.  This was a randomized, placebo-controlled
dose response trial for the treatment of acute migraine, with a total of 7
active doses ranging between 2.5mg and 200mg and placebo.  The primary
endpoint was ``being pain free at 2 hours postdose'', \textit{i.e.},  a
binary endpoint. The analysis presented here is a post-hoc analysis.

As a reasonable set of candidate models and contrasts, we select 4
different shapes of the sigmoid Emax model $f(x,\bm
\theta)=E_0+E_{max}x^h/(x^h+ED_{50}^h)$, which cover a very wide
range of monotonic shapes and a quadratic model to
safeguard against the possibility of a unimodal dose response
relationship. The \texttt{Mods} function is used for that, and one can
also plot the candidate shapes.

\begin{verbatim}
doses <- c(0,2.5,5,10,20,50,100,200)
models <- Mods(sigEmax = rbind(c(2.5, 1),c(10,1),c(50, 3),c(100,2)),
               quadratic = -1/250, doses=doses)
plot(models)
\end{verbatim}

The first stage of the two-step MCPMod approach consists of fitting a
model with ANOVA-type parametrization to the data to obtain estimates
$\widehat{\bmu}$ and its asymptotic covariance matrix $\bm S$. The
logistic regression model is used here, which means that the candidate
models are formulated on the logit scale (other scales could be used).
The ANOVA logistic regression model can be fitted as follows.
\begin{verbatim}
## data from NCT00712725 study
dosesFact <- as.factor(doses) ## treat dose as categorical variable
N <- c(133, 32, 44, 63, 63, 65, 59, 58)
## % of patients painfree at 2h post-dose
RespRate <- c(13,4,5,16,12,14,14,21)/N
## fit logistic regression (without intercept)
logfit <- glm(RespRate~dosesFact-1, family = binomial, weights = N)
muHat <- coef(logfit)
S <- vcov(logfit)
\end{verbatim}

Now all subsequent inference only depends on \texttt{muHat} and
\texttt{S} obtained from the logistic regression. The multiple
contrast test from \ref{ssec:impl} using optimal trend contrasts can
be produced as follows
\begin{verbatim}
MCTtest(doses, muHat, S=S, models = models, type = "general")
\end{verbatim}

All contrasts are significant. The modeling step can now be performed
using the \texttt{fitMod} function. Here for illustration we fit the
sigmoid Emax model and the quadratic model.

\begin{verbatim}
modSE <- fitMod(doses, muHat, S=S, model = "emax", type="general")
modQuad <- fitMod(doses, muHat, S=S, model = "quadratic", type = "general")
gAIC(modSE);gAIC(modQuad)
\end{verbatim}

A comparison of the gAIC values reveals that the sigmoid
Emax model provides a better fit than the quadratic
model.

Above we performed the different steps of the MCPMod procedure
separately. One could alternatively have used
\begin{verbatim}
MCPMod(doses, muHat, S=S, models=models, type = "general", Delta = 0.2)
\end{verbatim}
directly.

\subsubsection{Fitting on placebo-adjusted scale}
\label{sec:rel}

For this example we use the \texttt{IBScovars} data set from the
\texttt{DoseFinding} package, taken from \cite{bies:hoth:2002}. The
data are part of a dose ranging trial on a compound for the treatment
of irritable bowel syndrome with four active doses 1, 2, 3, 4
equally distributed in the dose range $[0,4]$ and placebo. The primary
endpoint was a baseline adjusted abdominal pain score with larger
values corresponding to a better treatment effect. In total 369
patients completed the study, with nearly balanced allocation across
the doses.

The endpoint is assumed to be normally distributed and it is of
interest to adjust for the additive covariate \texttt{gender}. While
the \texttt{DoseFinding} package can deal with this situation exactly,
here we illustrate using the placebo-adjusted (effect) estimates. Note
that, in the case of time-to-event data, one would proceed
similarly. Here we only illustrate fitting an emax model, using the
\texttt{MCTtest} and \texttt{MCPMod} functions is analogous to the
calls in Section \ref{sec:example}, but using the additional argument
\texttt{placAdj = TRUE}. We plot the fitted model together with
confidence intervals for the model fit and the ANOVA type effect
estimates.
\begin{verbatim}
data(IBScovars)
anovaMod <- lm(resp~factor(dose)+gender, data=IBScovars)
drFit <- coef(anovaMod)[2:5] # placebo adjusted (=effect) estimates at doses
S <- vcov(anovaMod)[2:5,2:5] # estimated covariance
dose <- sort(unique(IBScovars$dose))[-1] # vector of active doses
## now fit an emax model to these estimates
gfit <- fitMod(dose, drFit, S=S, model = "emax", placAdj = TRUE, 
               type = "general", bnds = c(0.01, 2))
plot(gfit, CI = TRUE, plotData = "meansCI")
\end{verbatim}

\section{Conclusions}
\label{sec:concl}
The extended MCPMod methodology, together with its corresponding
software implementation in the \texttt{DoseFinding} package in
\texttt{R}, greatly broaden the scope of application of the original
MCPMod approach. Most type of endpoints, and associated model-based
analyses, utilized in dose finding studies can be handled in the
context of the extended approach.

Further extensions of the approach discussed here are possible and of
interest in practice. An increasing number of indications and drugs
require regimen selection, in addition to the more traditional dose
selection. Different approaches can be considered in the context of
MCPMod, or its extension, discussed here. One could focus on
estimating the exposure-response relationship, for example, combing
dose and regimen into one model covariate. The much larger number of
exposure values, compared to dose levels, could pose a problem for the
derivation of optimal model contrasts and for the MCP step, more
generally. Dose-time response modeling in the context of model
uncertainty provide another venue for extending MCPMod. Further
research is needed in those topics.

Model-based dose finding methods, such as the extended MCPMod, provide
better understanding of the dose-response relationship, generally
translating into more accurate dose selection for confirmatory
trials. Realizing the full potential that these methods have to offer,
however, requires changes in the way Phase II studies are
traditionally designed. By and large, dose finding studies are planned
as mini Phase III trials, using hypothesis tests to select the dose,
or doses, to bring forward to the confirmatory stage. Relatively few
doses (typically two active treatment arms and placebo) are used in
such Phase II studies, making it hard to entertain any type of
modeling. The sample size derivation in this type of studies is based
on power calculations for detecting at least one dose significantly
different from placebo. The resulting number of subjects is typically
inadequate for proper estimation of target doses (and dose response
modeling). A discussion of a different balance in resource
allocation between Phases II and III, taking into account the overall
probability of program success, is long overdue. Utilization of larger
number of doses (e.g., 4 or 5), coupled with larger sample sizes,
would go along way in enabling model-based methods to improve dose
selection in Phase II and, as a result, the probability of success in
Phase III.

%% \textbf{Acknowledgements:}\\
%% We would like to thank several colleagues who have greatly contributed
%% to the development, extension, and application of MCPMod over the
%% years, in particular: Chyi-Hung Hsu, David Ohlssen, Georgina
%% Bermann, and Francois Mercier.

\appendix

\section{Derivation of Optimal Contrasts}
\label{sec:optcont} In this section the closed form solution for the
optimal contrasts for the case of a general covariance structure is
derived. Optimality here refers to maximum power of the univariate
contrast test, if a specified mean vector $\bm \mu$ (with
corresponding positive definite covariance matrix $\bm S$) is true,
which means that the non-centrality parameter
$$g(\mathbf{c},\boldsymbol{\mu})=\frac{\mathbf{c}'\bmu}{\sqrt{\bm
    c'\bm S\bm c}},$$ needs to be maximized with respect to $\bm c$,
subject to $\bm c'\bm 1=0$.

Writing $\bm C_0=\left(-\mathbf{1}_{K-1} \vdots
\mathbf{I}_{K-1}\right)$, the constrained maximization problem is
equivalent to the unconstrained maximization of
$\frac{\left(\mathbf{c}'\mathbf{C_0\mu}\right)^2}{\mathbf{c}'\mathbf{C_0SC_0}'\mathbf{c}}$. This,
however, is the solution to the generalized eigenvalue problem
\[
\bm C_0\bmu\bmu'\bm C_0'\bm x=\lambda \mathbf{C_0SC_0}' \bm x,
\]
see e.g. \cite{ahre:laeu:1981}, formula (2.66). As $\bm
C_0\bmu\bmu'\bm C_0'$ is of rank 1, it has only a single non-zero
eigenvalue. Thus, it is immediately clear that
$\mathbf{c}=const\cdot(\mathbf{\bm C_0S\bm C_0'})^{-1}\bm C_0\bmu,
const \neq 0$ is the only solution to the generalized eigenvalue
problem.  We further note that $\mathbf{c}=const\cdot(\mathbf{\bm
  C_0S\bm C_0'})^{-1}\bm C_0\bmu =const \cdot {\bm
  S}^{-1}(\bmu-\frac{\bmu'{\bm S}^{-1}\bm 1}{\bm1'{\bm
    S}^{-1}\bm1}\bm1)$.  It is clear that the optimal solution is
invariant with respect to addition or multiplication of any scalar
constant to the vector $\bm \mu$, which is why one can also use the
standardized mean vectors $\bm \mu^0_m$ instead of $\bmu$, which then
gives the result in formula (\ref{eq:optCont}).

\newpage

\section{Placebo-adjusted dose response modeling}
\label{ref:diff2plac}

In a few cases one would like to perform MCPMod on placebo-adjusted
estimates, for example when there are (additive) covariates in the
model, or when using a Cox PH model (where one can only obtain
control-adjusted estimates). In what follows we will first demonstrate
the equality of test statistics, and calculate optimal contrasts. Then

\subsection{Test statistics and optimal contrasts}
\label{sec:tt}

If we want to test a linear contrast of the responses per dose group,
it does not matter whether we fit a placebo adjusted curve or include
the placebo group as a response and then test contrasts to placebo.

To see this, consider the ANOVA estimate
\begin{equation}\label{eqA41}
\widehat{\bm \mu}\sim N(\bm \mu, \bm S)
\end{equation}
where the first component $\widehat{\mu}_0$ of the vector
$\widehat{\bm \mu}$ corresponds to the placebo response.  The test
contrasts can then be produced by multiplication of $\widehat{\bm
  \mu}$ with the $(K-1)\times K$ contrast matrix $\bm
C_0=\left(-\mathbf{1}_{K-1} \vdots \mathbf{I}_{K-1}\right)$, where
$\bm 1_{K-1}$ is a column vector of ones of size $K-1$ and
$\bm{I}_{K-1}$ the $K-1$ dimensional identity matrix. We obtain the
corresponding contrast
\begin{equation}\label{eqA42}
\widehat{\bm \mu}_C\sim N(\bm \mu_C, \bm S_C)
\end{equation}
with $\widehat{\bm \mu}_C=\bm C_0\widehat{\bm \mu}$, $\bm \mu_C=\bm C_0\bm \mu$ and
$\bm S_C= \bm C_0\bm S\bm C_0$.

The contrast test statistic in model (\ref{eqA41}) is of the form
\begin{equation} \label{f3}
m_C=\max
\frac{\mathbf{c}'\widehat{\bm \mu}}{\sqrt{\mathbf{c}'\mathbf{S}\mathbf{c}}}
\mbox{ subject to } \mathbf{c}'\mathbf{1}_K=0,
\end{equation}
with $\mathbf{c}$ such that $m_C$ attains a maximum.  In model
(\ref{eqA42}), the restriction on $\mathbf{c}$ is already absorbed in
the matrix $\bm C_0$ and the test statistic takes the form
\begin{equation} \label{f2}
m_P=\max
\frac{\mathbf{d}'\mathbf{C}_0 \bm \mu}{\sqrt{\mathbf{d}'\bm C_0 \mathbf{SC}_0'\mathbf{d}}},
\end{equation}
where $\mathbf{d}$ is no longer a contrast and again $\mathbf{d}$ is
chosen such that $m_P$ is at its maximum.  Now, it can be seen that
$\mathbf{c}=\bm C_0'\mathbf{d}$: Setting
$\mathbf{c}=\mathbf{C}_0'\mathbf{d}$ implies
$\mathbf{c}'\mathbf{1}_K=0$, as $\bm C_0\mathbf{1}_K=0$. Hence,
$m_C\geq m_P$. On the other hand, if $\mathbf{c}'\mathbf{1}_K=0$, then
there must be some $\mathbf{d}\in \mathbf{R}^{K-1}$ such that
$\mathbf{c}$ can be written as $\mathbf{c}=\mathbf{C}_0'\mathbf{d}$,
since the rows of the $(K-1)\times K$-matrix $\mathbf{C}_0$ provide a
base of the $(K-1)$-dimensional hyperspace orthogonal to
$\mathbf{1}_K$ in $\mathbf{R}^K$. Consequently, $m_C\leq m_P$. It
follows that $m_C = m_P$ and that if $\mathbf{d}$ maximizes
(\ref{f2}), then $\mathbf{C}_0'\mathbf{d}$ maximizes (\ref{f3}).

Specifically the optimal $\mathbf{d}^{opt}$ can be calculated as
$\mathbf{d}^{opt}=\bm S_C^{-1}\bm \mu_C$, as the sum to 0 constraint
is removed.

\subsection{Dose Response Model Fitting}
\label{sec:dose response-model}

In the two-stage generalized least squares fitting procedure one
minimizes the criterion
\begin{equation}
  \label{eq:drfituadj}
  (f(\bm x,\bm \theta)-\widehat{\bm \mu})'\bm S^{-1}(f(\bm x,\bm \theta)-\widehat{\bm \mu}).
\end{equation}

When we only have $\widehat{\bm \mu}_C$ one would not fit a full
dose response model $\theta_0+\theta_1 f^0(x,\bm \theta^0)$ but work
with a model without the intercept $\theta_0$.  $f_C(x,\bm \theta) =
\theta_1 f(x,\bm \theta^0)$. The optimization criterion proposed for
placebo-adjusted is then
\begin{equation}
  \label{eq:drfitadj}
  (f_C(x,\bm \theta)-\widehat{\bm \mu}_C)'\bm S_C^{-1}(f_C(x,\bm \theta)-\widehat{\bm \mu}_C).
\end{equation}

When $f^0(0, \bm \theta^0)=0$ one can see that (\ref{eq:drfituadj})
and (\ref{eq:drfitadj}) are equal. This follows from the fact that
(\ref{eq:drfitadj}) is equal to
\begin{equation*}
  (\bm C_0(f(\bm x,\bm \theta)-\widehat{\bm \mu}))'(\bm C_0\bm S\bm C_0')^{-1}(\bm C_0(f(\bm x,\bm \theta)-\widehat{\bm \mu})),
\end{equation*}

and because $\bm C_0'(\bm C_0\bm S\bm C_0')^{-1}\bm C_0 =\bm S^{-1}$
(which follows from multiplication from the left with $\bm C_0\bm S$).

\section{Proof of the result in section \ref{sec:asymp}}
\label{app:b}

Let $\bm \theta_0$ denote the true value of the parameter $\bm \theta$
and $\bm \mu_0 = \bm f(\bm x, \bm \theta_0)$ where $\bm x$ is a known
vector of fixed values.  Assume that the following conditions are
satisfied:
\begin{itemize}
\item[(A1)] There exists a symmetric, positive definite estimate $\bm
  S$ of the covariance matrix of $\widehat{\bm \mu}$ with $a_n\bm
  S \overset{P}{\rightarrow} \bm \Sigma$ for a positive
  nondecreasing sequence $a_n$ converging to $\infty$ as $n\rightarrow
  \infty$, and a positive definite, symmetric matrix $\bm \Sigma \in
  \mathbb{R}^{d\times d}$.
\item[(A2)] If $N(0, \bm \Sigma)$ denotes the multivariate normal
  distribution with mean 0 and covariance matrix $\bm \Sigma$, then
  ${a_n}^{1/2}(\widehat{\bm \mu}-\bm
  \mu_0)\overset{d}{\rightarrow}N(0, \bm \Sigma)$, where
  $\overset{d}{\rightarrow}$ denotes convergence in distribution. As a
  consequence of $a_n\rightarrow \infty$, the estimate $\widehat{\bm
    \mu}$ is consistent, \textit{i.e.}, $\widehat{\bm \mu}
  \overset{P}{\rightarrow}\bm \mu_0$ (see e.g. Serfling, 1980, p.26).
\item[(A3)] The mapping $\bm \Theta \mapsto \mathbb{R}^k$, $\bm \theta
  \mapsto \bm f(\bm x, \bm \theta)$, with $\bm x \in \mathbb{R}^k$ is
  a bijective function of $\bm \theta$ which is twice differentiable
  in an open region around $\bm \theta_0$.
\end{itemize}

Under these assumptions $\widehat{\bm \theta}$ is a consistent
estimator of $\bm \theta$, i.e., $\widehat{\bm \theta}
\overset{P}{\rightarrow} \bm \theta$ and $\sqrt{a_n}(\widehat{\bm
  \theta}-\bm \theta_0)\overset{d}{\rightarrow}N(\bm 0,\bm B(\bm
\theta_0)'\bm M(\bm \theta_0)\bm B(\bm \theta_0))$, where $\bm M(\bm
\theta)=\bm F(\bm \theta)'\bm A\bm \Sigma \bm A\bm F(\bm \theta)$ and
$\bm B(\bm \theta)=(\bm F(\bm \theta)'\bm A\bm F(\bm \theta))^{-1}$.

\textbf{\textit{Proof:}}

Let $\widehat{\bm \Psi}({\bm \theta})=(\widehat{\bm \mu}-\bm f(\bm x, \bm \theta))'
{\bm A_n}(\widehat{\bm \mu}-\bm f(\bm x, \bm \theta))$ and ${\bm \Psi}({\bm \theta})=({\bm \mu}-\bm f(\bm x, \bm \theta))'
{\bm A}({\bm \mu}-\bm f(\bm x, \bm \theta))$.

First we note that consistency of the estimator is easy to establish
using standard theory for the consistency of M-estimators. For example
the three conditions in Theorem 5.7 from \cite{vaar:1998} are easy to
verify (A1)-(A3).

The proof of the distribution of $\sqrt{a_n}(\widehat{\bm \theta}-\bm
\theta_0)\overset{d}{\rightarrow}N(\bm 0,\bm B(\bm \theta_0)'\bm M(\bm
\theta_0)\bm B(\bm \theta_0))$ works along the lines of
\cite[ch. 12.2.3]{sebe:wild:1989}, which we restate here with the
modifications needed for our situation.\\
As $\widehat{\bm \theta}$ minimizes $\widehat{\bm \Psi}(\bm \theta)$,
we have$\frac{d \widehat{\bm \Psi}(\widehat{\bm \theta})}{d \bm
  \theta}=0$. Thus, by the mean value theorem, there is a
$\widetilde{\bm \theta}$ between $\bm \theta_0$ and $\widehat{\bm
  \theta}$ such that $$0=\frac{d \widehat{\bm \Psi}(\bm \theta_0)}{d
  \bm \theta}+\frac{d^2\widehat{\bm \Psi}(\widetilde{\bm
    \theta)}}{d\bm \theta d\bm \theta'}(\bm \theta_0-\widehat{\bm
  \theta}).$$ Hence
\begin{equation}\label{eqA1}
\sqrt{a_n}(\widehat{\bm \theta}-\bm \theta_0)=\sqrt{a_n}(\frac{d^2\widehat{\bm
    \Psi}(\widetilde{\bm \theta)}}{d\bm \theta d\bm
  \theta'})^{-1}(\frac{d \widehat{\bm \Psi}(\bm \theta_0)}{d \bm
  \theta}).
  \end{equation}

We now show that (i) $\sqrt{a_n}\frac{d \widehat{\bm \Psi}(\bm
  \theta_0)}{d \bm \theta}$ is asymptotically normal, and that (ii)
$\left(\frac{d^2\widehat{\bm \Psi}(\widetilde{\bm \theta)}}{d\bm
  \theta d\bm \theta'}\right)^{-1}$ converges in probability to a
non-singular matrix.

(i)
$$\frac{d \widehat{\bm \Psi}({\bm
    \theta})}{d \bm \theta}= -2\bm F(\bm \theta)
\bm A_n(\widehat{\bm \mu}-\bm f(\bm x, \bm
\theta)),$$ where $\bm F(\bm \theta)$ is the $d \times k$ matrix of partial derivatives $\frac{d f(x_i,\bm \theta)}{d \theta_h}$. ($i=1,...,k$,$\quad
h=1,...,d$).

Since $\sqrt{a_n}\left(\widehat{\bm \mu}-\bm f(\bm x, \bm
\theta_0)\right)\overset{d}{\rightarrow}N(\bm 0,\bm
\Sigma)$, $$\sqrt{a_n}\frac{d \widehat{\bm \Psi}(\bm
  \theta_0)}{d \bm \theta}=-2\sqrt{a_n}\bm F(\bm \theta_0) \bm A_n\left(\widehat{\bm \mu}-\bm f(\bm x, \bm
\theta_0)\right)\overset{d}{\rightarrow}
N(\bm 0, 4\bm F(\bm \theta_0)\bm A\bm
\Sigma \bm A'\bm F(\bm \theta_0)')$$.

(ii) Differentiate the second term twice to get
$$\frac{d^2\widehat{\bm \Psi}(\widetilde{\bm
    \theta)}}{d\bm \theta d\bm \theta'}=-2(\bm U- \bm
F(\widetilde{\bm \theta})\bm A_n\bm F(\widetilde{\bm \theta})'),$$

where $\bm U$ is the $d\times d$ matrix with $h$-th column given by
$$\frac{d^2 f(\bm x, \bm
    \theta)}{d \theta_h d\bm \theta'}\bm A_n(\widehat{\bm \mu}-\bm f(\bm
x, \widetilde{\bm \theta)}).$$ Now $\widehat{\bm \mu}
\overset{P}{\rightarrow} \bm f(\bm x, \bm \theta_0)$ and
$\widehat{\bm \theta}\overset{P}{\rightarrow}\widetilde{\bm \theta}\overset{P}{\rightarrow}\bm \theta_0$, so all
entries of $\bm U$ converge to 0. In total we get
$$\frac{d^2\widehat{\bm \Psi}(\widetilde{\bm
    \theta)}}{d\bm \theta d\bm \theta'}\overset{P}{\rightarrow} 2\bm F(\bm
\theta_0)\bm A\bm F(\bm \theta_0)'$$

Defining $\bm M(\bm \theta)=\bm F(\bm \theta)\bm A\bm \Sigma \bm A'\bm
F(\bm \theta)'$ and $\bm B(\bm \theta)=(\bm F(\bm \theta)\bm A\bm
F(\bm \theta)')^{-1}$ and inserting the results from (i) and (ii) into (\ref{eqA1}), one obtains that the asymptotic distribution of
$\sqrt{a_n}(\widehat{\bm \theta}-\bm \theta_0)$ is $N(\bm 0, \bm B(\bm
\theta_0)'\bm M(\bm \theta_0)\bm B(\bm \theta_0))$. $\Box$

\section{Additional Plots}
\label{ref:addp}

\begin{figure}
  \begin{center}
  \includegraphics[width=0.7\textwidth]{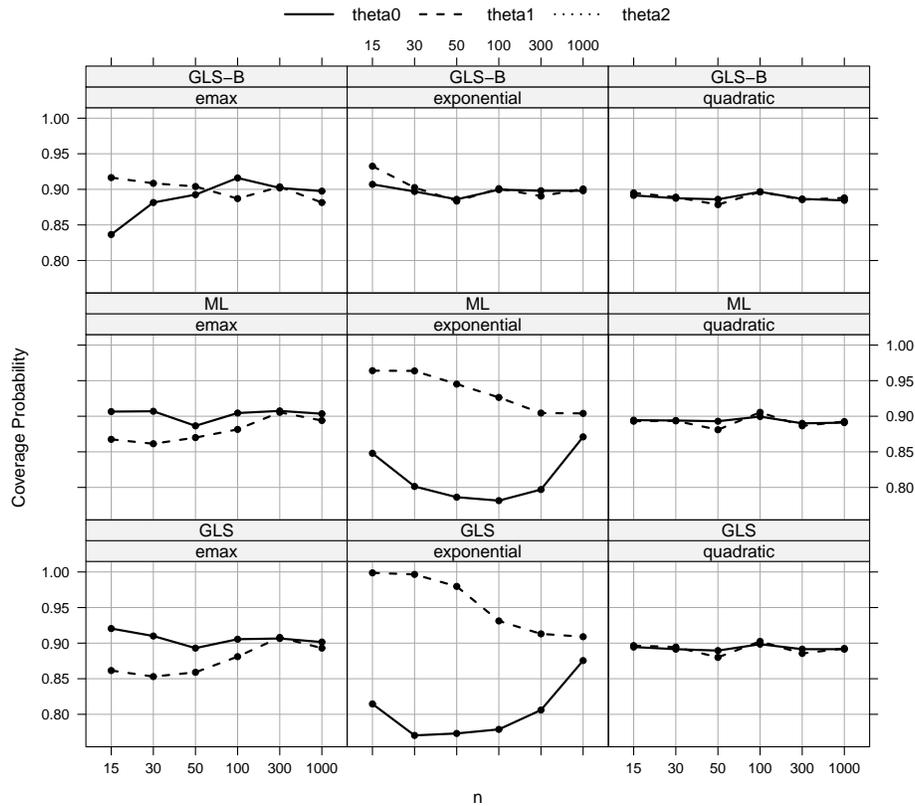}
  \end{center}
\caption{Time-To-Event endpoint}
\label{fig:CIsurv}
\end{figure}

\begin{figure}
  \begin{center}
  \includegraphics[width=0.7\textwidth]{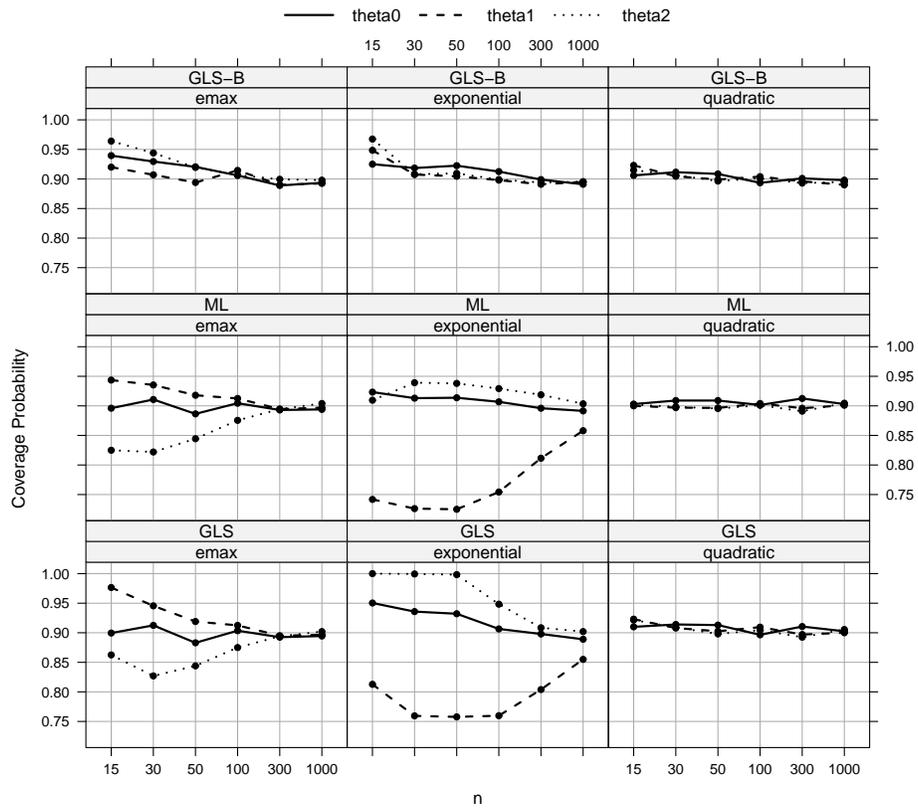}
  \end{center}
\caption{Binary endpoint}
\label{fig:CIbin}
\end{figure}

\begin{figure}
  \begin{center}
  \includegraphics[width=0.7\textwidth]{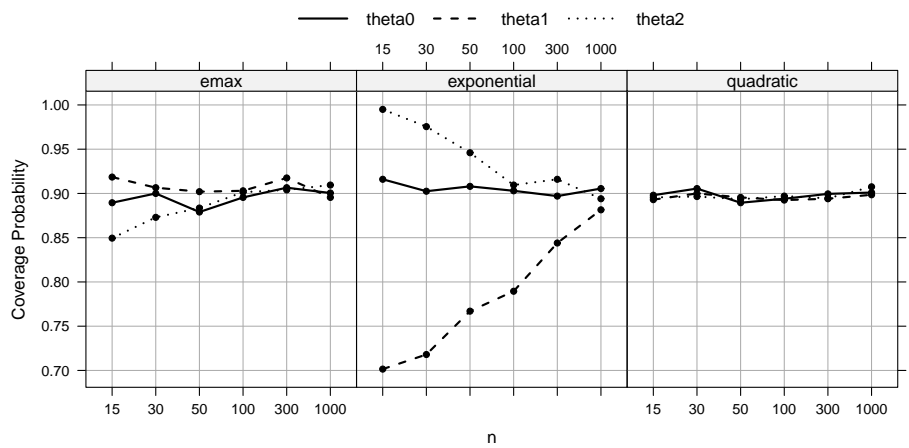}
  \end{center}
\caption{Normal endpoint}
\label{fig:CInorm}
\end{figure}

\begin{figure}
  \begin{center}
  \includegraphics[width=0.75\textwidth]{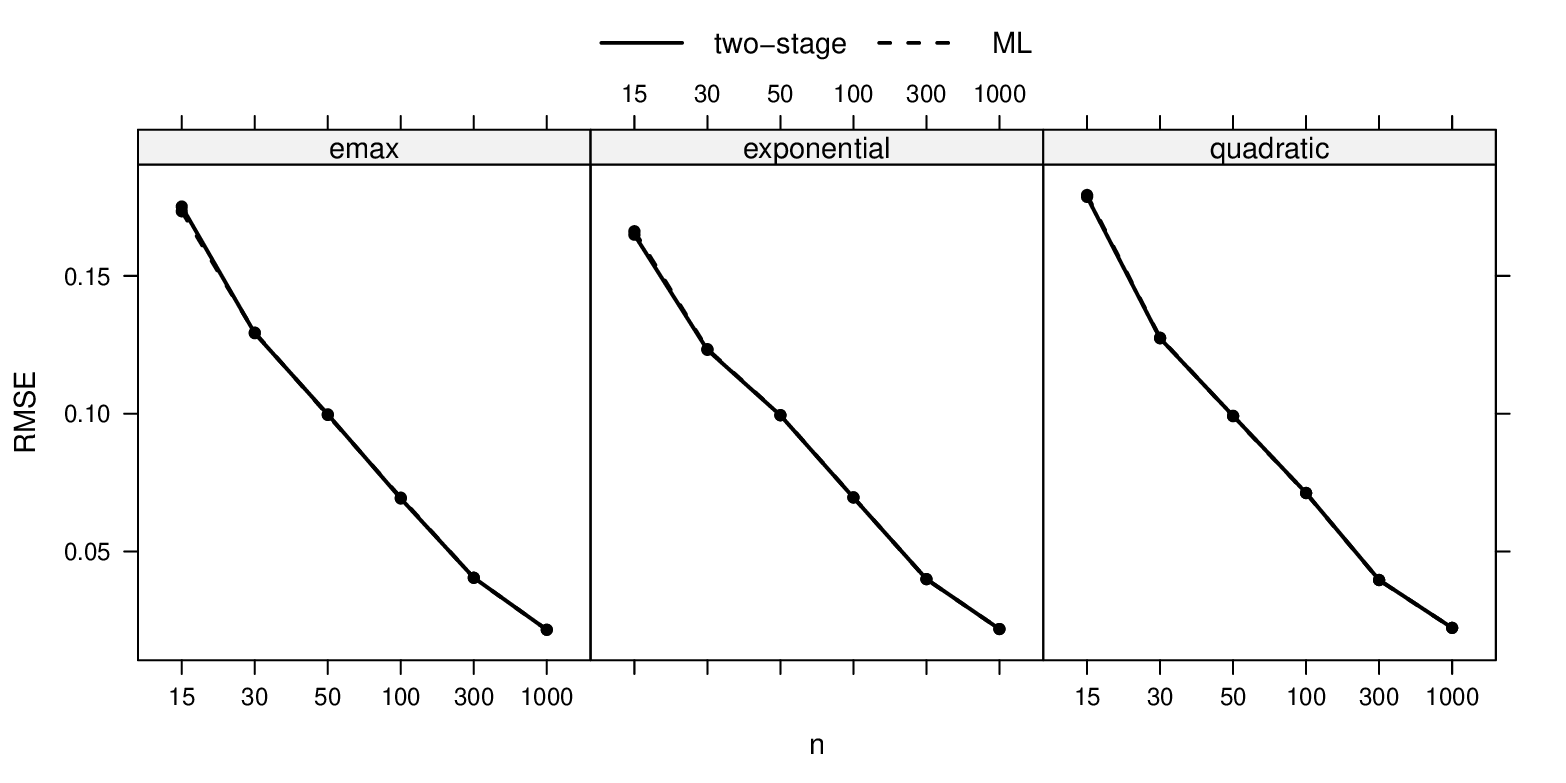}
  \end{center}
\caption{Root mean squared error for estimating the dose response
  (averaged over the doses available), count endpoint.}
\label{fig:drest}
\end{figure}

\begin{figure}
  \begin{center}
  \includegraphics[width=0.75\textwidth]{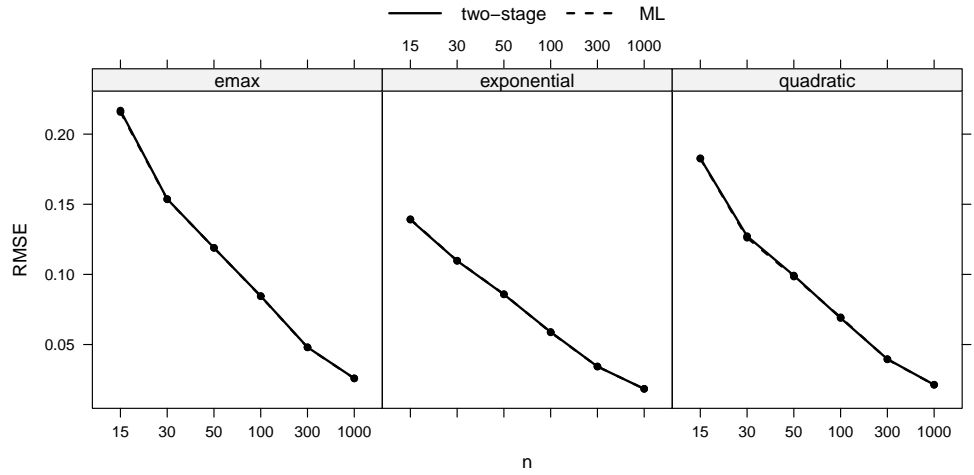}
  \end{center}
\caption{Root mean squared error for estimating the dose response
  (averaged over the doses available), time-to-event endpoint.}
\label{fig:estbin}
\end{figure}

\begin{figure}
  \begin{center}
  \includegraphics[width=0.75\textwidth]{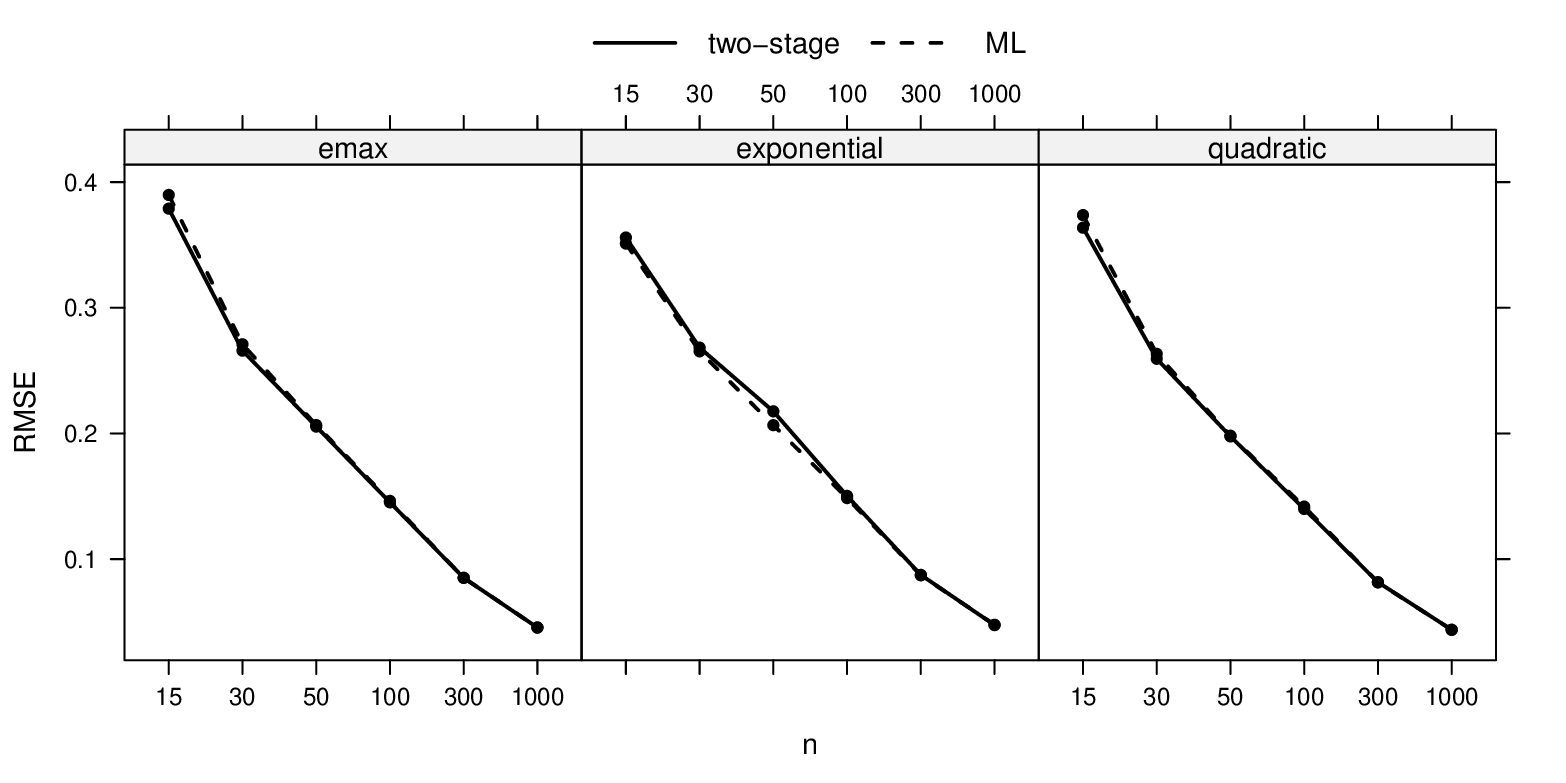}
  \end{center}
\caption{Root mean squared error for estimating the dose response
  (averaged over the doses available), binary endpoint.}
\label{fig:est}
\end{figure}

\bibliographystyle{plain}
\bibliography{bibl}

\end{document}